\documentclass[11pt]{article}
\usepackage[margin=1in]{geometry}
\usepackage{setspace}
\usepackage{amsmath,amssymb}
\usepackage{graphicx}
\usepackage{booktabs}
\usepackage{array}
\newcolumntype{P}[1]{>{\raggedright\arraybackslash}p{#1}}
\usepackage{caption}
\usepackage[authoryear,round]{natbib}
\PassOptionsToPackage{hyphens}{url}\usepackage{url}
\usepackage[colorlinks=true,linkcolor=blue,citecolor=blue,urlcolor=blue]{hyperref}
\newcommand{\tnote}[1]{\par\vspace{2pt}\begin{minipage}{\textwidth}\footnotesize\setstretch{1.05}\textit{Notes:} #1\end{minipage}}
\newcommand{\dom}[1]{\texttt{#1}}
\newcommand{\newapp}[1]{\setcounter{table}{0}\setcounter{figure}{0}\renewcommand{\thetable}{#1\arabic{table}}\renewcommand{\thefigure}{#1\arabic{figure}}}
\newcommand{\accessed}{accessed November 23, 2026}

\title{Free to Leave, Choosing to Stay:\\[4pt] Websites and Google's AI Overviews Opt-Out}
\author{Meng Liu\thanks{Olin Business School, Washington University in St.\ Louis. Email: mengl@wustl.edu.}}
\date{November 2026}

\begin{document}
\maketitle

\begin{abstract}
\noindent Until 2026, a website could keep its content out of Google's AI Overviews only by giving up search traffic. Publishers and regulators called this a tie. After a UK competition order, Google added a setting, open to every site by August 31, 2026, that removes a site from AI Overviews with no penalty in ordinary search. We provide some of the earliest evidence on whether websites opted out in the first eleven weeks. Combining two data sets on Google search results pages and the AI Overviews on them, we report three pieces of evidence that very few sites did. First, at the end of the first week, more than 99.5 percent of sites, including those that block AI crawlers, were still being cited in AI Overviews. Second, by mid-November the share of sites that go uncited in AI Overviews had not risen, across website types and sizes. Third, a difference-in-differences design compares sites that block AI crawlers, which have more reason to use the opt-out, with sites that block none, and finds that the blockers' citation rate did not fall relative to the comparison group after the change. Overall, removing the search penalty led to almost no detectable withdrawal from AI Overviews in the short run. These findings suggest that leaving AI Overviews is costly for most individual websites, and that the tie did not bind most of them to begin with.
\end{abstract}


\newpage

\section{Introduction}

Google's AI Overview is a generated answer that appears above the search results and lists the web pages it drew on. By mid-2026 it appeared on about 40 percent of Google searches in the United States.\footnote{Similarweb reports an AI Overview on 43 percent of US searches, up from 15 percent a year earlier (\emph{TechCrunch}, July 27, 2026, \url{https://techcrunch.com/2026/07/27/googles-ai-search-is-rapidly-becoming-the-default-new-data-shows/}, \accessed).} Two randomized experiments find that it cuts clicks to websites by about 40 percent on the searches where it appears \citep{agarwalsen2026,wang2026}.

Individual publishers object to more than the lost clicks. Their complaint is that they could not refuse.\footnote{For example, DMG Media, publisher of the Daily Mail, told the UK Competition and Markets Authority in 2025 that a publisher could not keep its content out of AI Overviews without also giving up crawling for Search, which it called ``not commercially viable'' (\url{https://assets.publishing.service.gov.uk/media/68a5aa50a6acbbc7fb96a3b2/DMG_Media.pdf}, paragraph 45). The chief executive of People Inc.\ made the same complaint in public (\emph{TechCrunch}, September 12, 2025, \url{https://techcrunch.com/2025/09/12/google-is-a-bad-actor-says-people-ceo-accusing-the-company-of-stealing-content}), as did the Independent Publishers Alliance, which has 45 members, in a complaint to the European Commission dated June 30, 2025 (\emph{Digiday}, July 8, 2025, \url{https://digiday.com/media/generative-ai-not-ad-tech-is-the-new-antitrust-battleground-for-google}) (all accessed November 26, 2026).} A website can turn away the crawlers that collect text to train AI models, Google's own included, with one line in its robots.txt file, and its place in Google Search is unaffected. The AI Overview was the exception: until 2026 a site could keep its content out of it only by giving up search traffic. Chegg and Penske Media sued Google on this ground in 2025,\footnote{A US district court dismissed both suits in 2026, and the plaintiffs have appealed (\emph{Press Gazette}, October 1, 2026, \url{https://pressgazette.co.uk/news/penske-ai-overviews-lawsuit-dismissed-because-no-formal-bargain-struck-with-google/}, \accessed).} the European Commission opened an investigation,\footnote{The European Commission's investigation, opened on December 9, 2025 (case AT.40983), asks whether Google used publishers' content for AI Overviews ``without offering them the possibility to refuse such use of their content'' (quoted in \emph{Press Gazette}, \url{https://pressgazette.co.uk/?p=246507}, accessed November 26, 2026).} and the UK Competition and Markets Authority (CMA) ordered Google to offer a way out. Some scholars call the arrangement a tie: to be listed in Search, a site had to supply its content to the overview \citep{singhscottmorton2026}.

What is a tie, and why does it matter? In a tie, a firm supplies something its trading partners cannot do without only if they also accept a second deal, and so obtains that deal on terms they would refuse if it were offered separately \citep{whinston1990,carltonwaldman2002}. Here the firm is Google, the trading partners are websites, what they cannot do without is a place in Google Search, and the second deal is supplying their content to AI Overviews. If the tie makes publishers supply content they would otherwise withhold or charge for, it harms two groups. Publishers earn less from the content they produce, and so have less reason to create it. Rival AI answer engines, such as those of OpenAI and Anthropic, cannot offer comparable search traffic in exchange, so they must pay for content that Google gets free. A tie can also save costs: Google uses one index for both the search results and the overview, and a single condition spares it from negotiating with millions of sites.

The claim against Google therefore has four parts. (1) Websites cannot do without Google Search. (2) Google made a place in Search conditional on supplying content to the overview. (3) The condition kept sites in the overview that would otherwise have left. (4) As a result, publishers were paid less, or rival AI answer engines were put at a disadvantage. Parts 1 and 2 are on record: a US court found in 2024 that Google holds a monopoly in general search, and Table~\ref{tab:controls} documents the condition in part 2. Part 4 depends on prices and deals that we do not observe. That leaves part 3, which this paper tests. The regulators' remedy rests on it as well, because an opt-out matters only if some sites use it. We say the tie \emph{binds} a site if the site would withdraw once withdrawing carries no search penalty. Before the opt-out existed this could not be tested, because every site faced the same penalty.

Then Google removed the penalty. A new setting, open to all websites worldwide from August 31, 2026, excludes a site from Google's generative search features, AI Overviews included, and Google states that it is not a ranking signal, so using it does not affect a site's position in ordinary Search. We call the setting \emph{penalty-free} in this narrow sense. It is not costless: a site that uses it gives up the clicks that its citations in AI Overviews bring, and the cost may be higher still if its competitors stay.

We study the sites with the clearest record of refusing AI: 22,000 sites whose robots.txt files block one or more AI crawlers, including Google-Extended, the control that keeps a site's content out of the training of Google's Gemini models. We compare them with 180,000 sites that block none, on archived Google results pages for 129,000 keywords, each captured twice before August 31 and once after. Only a site's owner can see the setting, so we infer a withdrawal from what we can observe: the site keeps its rank in the ordinary results but stops being cited in the AI Overview. A site can also stop being cited for reasons other than an opt-out, so the exits we count, if anything, overstate the use of the setting.

We report three findings. First, almost no site left in the first week. A second data source records each site's 50 most recent citations in AI Overviews, so a site that opted out would show a latest citation on the day it left. A week after the setting opened, more than 99.5 percent of 14,871 AI-blocking sites were still being cited, as were non-blocking sites. Second, there is no sign of exits through mid-November. On the results pages, the share of sites that rank in the top ten but are never cited in AI Overview shows no upward shift from the two five-week periods before August 31 to the two after. This holds for sites that block Google's AI training crawler, sites that block other AI crawlers and sites that block none, and for sites that rank on a single results page as well as those that rank on twenty or more.

Third, a difference-in-differences (DiD) style design shows that AI-blocking sites were cited as often as before. It follows the same site on the same keyword before and after August 31 and compares AI-blockers, the sites with the most reason to opt out, with non-blockers on the same results page. With very large sites capped in weight, sites that block Google-Extended, Google's AI training crawler, were cited 1.10 percentage points more often after August 31 (standard error 0.33). Against a baseline citation rate of 43 percent, this is an increase of 2.6 percent. Without the cap the estimate is negative, and it becomes small and insignificant once the 10 largest sites are removed. Among publishers and news sites alone, the estimates are small and mostly positive.

These results speak to how Google and content providers interact in generative search. Sites that had refused to supply training data stayed in the AI Overview once leaving was free, so their presence reflects a choice, not only a condition that Google imposed. A likely reason is that the two uses differ: AI training may return nothing or little to a site, while an AI Overview citation returns visibility and some clicks, and a site that leaves hands its place to a competitor. The same logic applies to other answer engines, such as ChatGPT and Perplexity, that send some traffic back. For policy, a right to opt out, the remedy the CMA adopted and publishers have sought from the European Commission, does not by itself change who supplies content to AI answers. If the concern is what publishers are paid, the remedy is more likely to lie in the terms of use, such as payment, attribution and collective bargaining, than in each site's right to leave. Whether that right strengthens publishers' bargaining position over time remains open.

\subsection{Related literature}\label{sec:lit}

\emph{Tying.} Theory shows when a tie excludes: it can deter entry into the tied market \citep{whinston1990} or protect the monopoly in the tying market \citep{carltonwaldman2002}, and when a platform cannot charge one side a price below zero, a tie lets it capture surplus it could not otherwise reach \citep{choijeon2021}. \citet{singhscottmorton2026} apply these ideas to AI Overviews, where the tie is on the buying side. \citet{cohendavies2026} predicted that publishers would not use an opt-out, because they depend on search traffic. Evidence on whether a tie changes what the tied party does is scarce, because ties are seldom switched off. We observe one being switched off, and so far their prediction holds.

\noindent \emph{Platforms that reuse publishers' content.} News aggregators send readers to publishers and also substitute for them \citep{jeonnasr2016}. When Google News dropped a publisher's content, closed in Spain, or stopped showing the snippets of German publishers that withheld consent, news sites lost visits \citep{chioutucker2017,athey2021,calzadagil2020}. Recent studies find that AI Overviews likewise reduce visits to the sites they draw on \citep{agarwalsen2026,wang2026,khosravi2026}, though they raised activity in the Reddit communities they can cite \citep{zhangcui2026}. AI chatbots refer readers even less often: a referral accompanies 5.4 percent of ChatGPT sessions, against 31.1 percent of Google queries \citep{shi2026}. Generative AI also competes with the creators it draws on: when a stock image marketplace admitted AI-generated images, sales of other images fell by 28 percent \citep{goldberglam2026}. This work measures what platforms and AI products do to those who supply their content. We measure what the suppliers do when they can refuse.

\noindent \emph{Websites' control over AI.} A growing share of websites restrict AI crawlers in robots.txt \citep{longpre2024,liu2025}, and the sites that do so are selected: among news sites, reputable and high-factual outlets block AI crawlers several times as often as misinformation and low-factual ones \citep{steinacker2026,zhu2026}. Creators withdraw as well: photographers whose images were placed in an AI training data set left the platform at a higher rate and uploaded less \citep{peukert2024}. \citet{zhaoberman2026} study how news publishers respond to generative AI, blocking included, and \citet{lin2026} finds that news sites lose traffic rank when they block OpenAI's search crawler but not its training crawler. Most of this work concerns the collection of training data, which a site can refuse at no cost. We study the refusal of a product that sends the site visitors. 

\section{Background: What a Website Could Refuse, and at What Cost}\label{sec:setting}

Google builds its AI Overview from pages in its search index and lists them as sources. Table~\ref{tab:controls} lists the four controls a site has had over this, and what each costs in ordinary Search.

\begin{table}[htbp]
\centering\small
\caption{What a Website Could Refuse, and at What Cost}\label{tab:controls}
\begin{tabular}{@{}P{4.3cm}P{3.2cm}P{2.9cm}P{4.4cm}@{}}
\toprule
Control & Available & Keeps the site out of AI Overviews & Cost in ordinary Google Search \\
\midrule
1. Block Googlebot in robots.txt, or \texttt{noindex}\textsuperscript{\ref{fn:c12}} & Always & Yes & The site's pages leave Search, or stay listed with no description \\
\hline
2. \texttt{nosnippet} or \texttt{max-snippet} tag\textsuperscript{\ref{fn:c12}} & Always & Yes & Listings lose their text snippet, which costs clicks \\
\hline
3. Block Google-Extended in robots.txt\textsuperscript{\ref{fn:c3}} & September 2023 & No & None \\
\hline
4. Search Console generative AI setting\textsuperscript{\ref{fn:c4}} & All websites worldwide from August 31, 2026; some UK owners from June 17 & Yes, and out of AI Mode and AI features in Discover & None stated: Google says it is not a ranking signal and therefore would not affect the ordinary Search \\
\bottomrule
\end{tabular}
\end{table}

Before 2026, the only controls that kept a site out of the overview, 1 and 2, both cost it search traffic.\footnote{\label{fn:c12}Google Search Central, ``AI features and your website,'' \url{https://developers.google.com/search/docs/appearance/ai-features}; ``Robots meta tag, data-nosnippet, and X-Robots-Tag specifications,'' \url{https://developers.google.com/search/docs/crawling-indexing/robots-meta-tag} (both \accessed).} Control 2 strips the description from the site's ordinary listings; the CMA cites a 2019 Google experiment in which a comparable display cut traffic to news publishers by 45 percent.\footnote{CMA, ``Consultation: Publisher Conduct Requirement,'' January 28, 2026, paragraph 1.11, \url{https://assets.publishing.service.gov.uk/media/6979d0bf75d44370965520a0/Publisher_conduct_requirement.pdf} (\accessed).} Few sites took this route: 0.2 percent of the listings in our data show no snippet (Online Appendix B.3).

Control 3, Google-Extended, was introduced in September 2023\footnote{\label{fn:c3}Google, ``An update on web publisher controls,'' September 28, 2023, \url{https://blog.google/innovation-and-ai/products/an-update-on-web-publisher-controls/} (\accessed).} and keeps a site's content out of the training of Google's Gemini models. It does not affect Search, and it does not keep a site out of the overview: sites that block it were cited on 40 percent of the overview pages where they ranked in the top ten before August 31, against 47 percent for sites that block nothing.

Control 4 is the subject of this paper. On June 3, 2026, the CMA required Google to let publishers withhold their content from generative AI features without being penalized in search, and Google announced a setting in Search Console, the tool through which owners manage their sites on Google.\footnote{\label{fn:c4}Google, ``New opportunities, control and insights for website owners,'' June 3, 2026, updated August 31, 2026, \url{https://blog.google/products-and-platforms/products/search/new-controls-website-owners/}; Search Console Help, ``Search generative AI control,'' \url{https://support.google.com/webmasters/answer/16908024}; CMA, ``Google search publisher conduct requirement,'' June 3, 2026, \url{https://www.gov.uk/find-digital-markets-measures/google-search-publisher-conduct-requirement} (all \accessed).} A site that turns it on is excluded from AI Overviews, AI Mode and the generative AI features in Discover together. Google states that exclusion takes effect within one to two days, that the site then receives no traffic from those features, and that the setting is not a ranking signal elsewhere in Search. The setting belongs to a Search Console \emph{property}, which can be a domain, a subdomain or a directory.

Google honored the setting for a subset of UK website owners from June 17, and on August 31 it opened the setting to all websites worldwide, US sites included. We treat August 31 as the date of the change and examine UK sites separately (Online Appendix B.5).

Did the setting work as described? Zyppy, a firm that advises websites on search engine optimization (SEO), excluded its own site from September 1 to 29 and reports that its impressions in AI features fell by more than 99 percent within an hour.\footnote{C.~Shepard, ``We Opted Out of Google AI Search: Here's What Happened,'' Zyppy, November 5, 2026, \url{https://signal.zyppy.com/p/we-opted-out-of-google-ai-search} (\accessed).} Our data show the same break: Zyppy is cited on 6 of the 10 overview pages where it ranked in the top ten shortly before August 31, and on none of 7 during its opt-out (Online Appendix C.2). Zyppy is the only site whose owner has publicly said it used the setting, so it is our one direct check that an opt-out looks in our data the way we assume: the site keeps ranking in Search and is no longer cited in AI Overview.

Only a site's owner can see the setting, so throughout we infer withdrawal from that pattern. A site that is still cited cannot have excluded its whole site. A trial that ends within weeks, as Zyppy's did, can fall between captures.

\section{Data}\label{sec:data}

This section describes the various data sources, key features, the construction of site types and the sample. A more detailed description of the data is offered in Online Appendix A.

\textbf{Results pages.} DataForSEO, a commercial provider of search data, archives Google results pages for the United States, capturing each keyword every six to ten weeks. A capture records the ranked organic results and the sources of the AI Overview, if there is one. Our keywords are a sample of 1,045,665 from the 10.1 million on which a second provider, Serpstat, recorded an AI Overview in August or September 2026; the sample favors keywords with high search volume (Online Appendix A.1), with a year of history for each.

A \emph{page} is one capture of one keyword on which an overview lists at least one source. We keep the keywords with at least two pages before August 31 and at least one after, and for each we use three pages: the last two before and the first after. This gives 129,057 keywords and 387,171 pages. The page after was captured in the first week for 29,842 keywords, later in September for 40,152, and between November 2 and 19 for nearly all of the other 59,063.

\textbf{Units.} A \emph{domain} is a registered name such as \dom{nytimes.com}. A \emph{site}, our unit of analysis, is a domain or a subdomain that serves its own robots.txt file, such as \dom{cooking.nytimes.com}; the sample has 202,335. A property, the unit of Google's setting, can be narrower than a site, and a publisher can own many sites; we do not group sites by owner. On each page we take the sites ranked in the top ten organic positions. A site is \emph{cited} if the overview lists any page from it, which happens 44 percent of the time.

\textbf{AI blockers.} We classify each site by its robots.txt file before August 31. The files come from Common Crawl's crawl of August 2026 and, for large sites missing from it, from the Internet Archive. A site \emph{blocks} a crawler if the file disallows the whole site for that crawler, or for all crawlers when the crawler is not named. \emph{Google-Extended blockers} block Google-Extended. \emph{Other-crawler blockers} block at least one of sixteen other AI crawlers, such as OpenAI's GPTBot, but not Google-Extended. \emph{Non-blockers} have a robots.txt file and block none of the seventeen. Sites with no archived file, 16 percent of observations, are left out. The sample has 14,638 Google-Extended blockers, 7,251 other-crawler blockers and 180,446 non-blockers, with 3.0 million site-by-page observations. About 11,000 of the blockers are publishers or news sites.

Blocking an AI crawler is an imperfect proxy for wanting out of the overview. Declining to train a model differs from declining a citation. And nearly two thirds of the Google-Extended blockers, 9,200 sites, carry the block through Cloudflare's managed robots.txt, which disallows a list of AI crawlers in one click. The other 5,400 wrote the rule themselves; they include nearly all the large sites. Table~\ref{tab:main} reports hand-written blocks separately.

\textbf{First-week records.} We supplement the DataForSEO data with a second source, Serpstat. Serpstat keeps, for each site, its 50 most recent citations in AI Overviews, with timestamps. Our copy ends on September 7, 2026. A site that stops being cited stays in the file with an older latest date. Of the sites with a full record of 50, 14,871 are AI blockers.

\section{Model-Free Evidence}\label{sec:facts}

\begin{figure}[!htb]
\centering
\includegraphics[width=\textwidth]{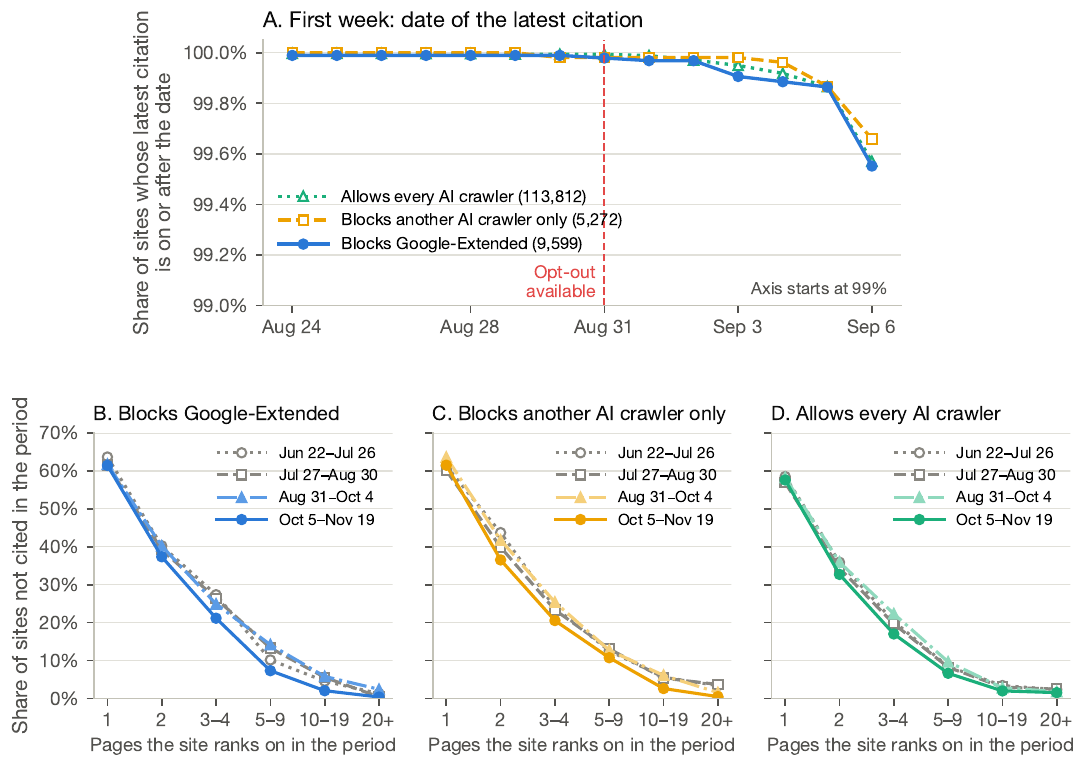}
\caption{Citations after the opt-out, by group of sites}\label{fig:exits}
\tnote{Panel A: sites with a full record of 50 citations in Serpstat; the records end on September 7. Panels B to D: the sites of each group that rank in the top ten on at least one page in a period; the two lines with open markers are the periods before August 31 and the two with filled markers the periods after (counts in Online Appendix Table~\ref{tab:bins}).}
\end{figure}

We do not observe which sites turned the setting on. We can only infer an opt-out from citations in AI Overviews: a site that opts out keeps its rank in the ordinary search results but is no longer cited. This section presents model-free evidence on how many sites left, as inferred from citations. Figure~\ref{fig:exits} draws on two data sources, each with its own limits.

We report everything for three groups of sites: Google-Extended blockers, other-crawler blockers and non-blockers. The reason, which also underlies the design of Section~\ref{sec:design}, is that the opt-out reached every site on the same day, so all sites were treated at the same time. What differs across sites is how likely they are to use it, which we infer from whether they refused AI crawlers before the change. Google-Extended blockers had already refused Google's AI where refusing cost nothing. If the tie held any sites in the overview against their wishes, these sites are the most likely to be among them, and so the most likely to opt out once the penalty was gone.

\emph{The first week.} We begin with the Serpstat records, which list each site's 50 most recent citations up to September 7. Their advantage is that they give the date of each site's latest citation. A large site is cited many times a day, so if it opted out, its latest citation would stay on the day it left, and the records would show that day. A small site is cited less often, and its latest citation can be several days old without any opt-out. We therefore use the sites with a full record of 50. Even among them an early latest date need not be an opt-out, so this measure, if anything, overstates opt-outs. It also covers only the first week.

Panel A plots, for each date, the share of sites whose latest citation falls on or after that date. If sites had opted out when the setting opened, the lines would fall soon after August 31. They barely move. In all three groups, about 99.9 percent of sites have their latest citation on September 5 or later, and at least 99.5 percent on September 6 or later (the vertical axis starts at 99 percent). At least 99.5 percent of the sites in each group therefore did not opt out in the first week. This is a lower bound, because some of the remaining sites had simply not been cited again when the records end (Online Appendix C.1 gives the full distribution). The one confirmed opt-out looks different in the same records: Zyppy's latest citation is on September 1, the day it turned the setting on.

\emph{Through mid-November.} Next we turn to the results pages, which cover the months before the change and eleven weeks after it. For each site we observe the pages on which it ranks in the top ten and whether the overview on each page cites it. We split the capture dates into four periods of five to seven weeks: June 22 to July 26 and July 27 to August 30 before the change, August 31 to October 4 and October 5 to November 19 after. For Google-Extended blockers, panel B plots, in each period, the share of sites that were not cited on any page against the number of pages on which the site ranked in that period. Plotting the data this way serves two purposes. First, opt-outs would be visible directly: a site that opted out is not cited, so the two lines for the periods after August 31 would lie above the two lines for the periods before. Second, the number of pages a site ranks on sorts sites by size, so we can look for opt-outs separately among small and large sites. Small sites need this care because they often go uncited by chance: most sites that rank on a single page are not cited, in any period.

The lines do not shift up after August 31. Among Google-Extended blockers that ranked on one page, 62 to 64 percent were not cited in the two periods before and 62 percent in each period after. Among those that ranked on 20 or more pages, the share was 1.3 and 0.4 percent before and 2.4 and 0.4 percent after. The 2.4 percent is eight sites; each had been cited on at most 15 percent of its pages before August 31, and six were cited again in the later period.

Panels C and D plot the same shares for other-crawler blockers and non-blockers, and the pattern is the same. Among other-crawler blockers that ranked on one page, 60 to 61 percent were not cited before and 62 to 64 percent after; among those that ranked on 20 or more, the share fell from 3.4 and 3.7 percent to 1.6 and 0.5. Among non-blockers, the share was 57 to 59 percent at one page in every period, and at 20 or more pages it was 2.3 and 2.5 percent before against 1.8 and 1.5 after. The two publishers that sued Google, Penske Media and Chegg, are still cited as often as before (Online Appendix B.7). Online Appendix B.6 follows the large sites one at a time.

\section{Difference-in-Differences}\label{sec:design}

\subsection{Design}\label{sec:did}

The model-free evidence suggests no substantial withdrawal across websites in the short run. At least 995 of every 1,000 sites were still being cited at the end of the first week, and through mid-November the share of sites that go uncited did not rise, among small sites or large ones. These comparisons give a simple and striking overall pattern: sites did not opt out in the weeks after the change. They leave three gaps, however, which motivate a difference-in-differences (DiD) design that follows the same site on the same keyword across successive results pages.

\emph{1. Blockers and non-blockers are not alike.} The keyword and the site's rank affect the probability of a citation, and they differ between the two groups. Even at the same rank on the same page, blockers were cited less often than non-blockers before the change (Online Appendix B.8). The model-free evidence lumps these differences together; the DiD design controls for them systematically.

\emph{2. Small exits are hard to detect in raw counts.} Panel B of Figure~\ref{fig:exits} shows that, among blockers that rank on one page, 62 percent are not cited in any given period. Had 3 percent of them opted out, the share would be 63 percent,\footnote{Of 100 such sites, 62 are not cited and 38 are. If 3 percent opted out at random, the opt-outs among the 62 would change nothing, and $0.03 \times 38 \approx 1$ of the 38 cited sites would become uncited. The share not cited would rise from 62 to about 63 percent.} yet the share moves by two points between the two periods before August 31, which may reflect trends or changes in the mix of keywords over time. The regression follows the same site on the same keyword, which removes that noise and is better suited to detecting small effects.

\emph{3. A site can leave in part.} An owner can exclude one directory, such as \dom{example.com/recipes/}, and keep the rest. The model-free evidence captures only the extensive margin at the site level, because a site that opts out in part may still be cited through its other pages. By tracking how often a site is cited on the same keyword over time, the DiD design captures both the extensive and the intensive margin.

\noindent We therefore estimate
\begin{equation}
\text{Cited}_{ikt} = \alpha_{kt} + \delta_{ik} + \gamma_{r(i,k,t)} + \beta_1\,\text{GExt}_i \times \text{After}_t + \beta_2\,\text{Other}_i \times \text{After}_t + \varepsilon_{ikt},
\label{eq:page}
\end{equation}
where $i$ is a site, $k$ a keyword and $t$ one of the keyword's three pages, which were captured on different dates. The page effects $\alpha_{kt}$ compare sites that appear together on one page, which removes whatever Google changed about that overview. The site-by-keyword effects $\delta_{ik}$ absorb any unobserved difference at the site-keyword level, so that we follow the same site on the same keyword over time. The rank effects $\gamma_{r(i,k,t)}$ allow citation to depend on the site's organic position on that page. $\text{GExt}_i$ indicates Google-Extended blockers, the sites that blocked Google-Extended before August 31, and $\text{Other}_i$ indicates the sites that blocked another AI crawler but not Google-Extended. $\text{After}_t$ indicates the pages captured on or after August 31. Standard errors are clustered by domain.

\textbf{What $\beta_1$ and $\beta_2$ measure.} Every site was offered the option from August 31, so $\beta_1$ captures the change in Google-Extended blockers' citation rate after August 31 relative to the change for non-blockers, and any \emph{additional} opt-out by the blockers would make it negative; $\beta_2$ does the same for other-crawler blockers. The design is in effect a DiD in exposure to the treatment. It assumes (1) that Google-Extended blockers and other-crawler blockers are more likely to use the option than non-blockers, and (2) that without the option the citation rates of blockers and non-blockers would have moved in parallel. Each keyword has only two pages before the change, so we cannot trace leads and lags keyword by keyword as in a standard DiD. We examine assumption (2) in two other ways: a placebo test between the two pages before, which finds differences that are zero or small and weakly significant (Table~\ref{tab:main}, panel B), and estimates by month of capture (Figure~\ref{fig:months}).

\textbf{Two weightings.} The ten largest sites hold 26 percent of observations, so we report equation~(\ref{eq:page}) two ways. The \emph{uncapped} estimate gives every observation the same weight; it describes where citations go, and the largest sites drive it. The \emph{capped} estimate gives an observation of a domain with $n>50$ observations the weight $50/n$, so that no domain counts for more than 50 observations; it describes the typical site, whose decision the tying claim is about.

\subsection{Results}

\begin{table}[htbp]
\centering\footnotesize\setlength{\tabcolsep}{2.6pt}
\caption{Change in citation rate after August 31, relative to non-blocking sites}\label{tab:main}
\begin{tabular}{@{}lccccr@{}}
\toprule
 & \multicolumn{2}{c}{Capped at 50 results per domain} & \multicolumn{2}{c}{Uncapped} & \\
\cmidrule(lr){2-3}\cmidrule(lr){4-5}
 & Blocks Google- & Blocks another & Blocks Google- & Blocks another & \\
 & Extended & AI crawler only & Extended & AI crawler only & Observations \\
\midrule
\multicolumn{6}{@{}l}{\emph{Panel A: Main results}} \\
All sites & 1.10*** (0.33) & 0.55 (0.41) & $-$2.60*** (0.91) & $-$1.05** (0.49) & 2,416,502 \\
Without the 10 largest sites & 1.14*** (0.33) & 0.58 (0.41) & 0.56 (0.46) & $-$0.29 (0.63) & 1,722,475 \\
Publishers and news sites & 1.00* (0.51) & $-$0.30 (0.63) & 1.00** (0.43) & 0.74 (0.46) & 513,017 \\
\quad with a hand-written block & 0.24 (0.66) & $-$0.32 (0.63) & 0.74 (0.50) & 0.73 (0.46) & 513,017 \\[3pt]
\multicolumn{6}{@{}l}{\emph{Panel B: All sites, between the two pages before August 31}} \\
Later page relative to earlier & 0.60* (0.36) & 0.76* (0.42) & 0.06 (0.69) & $-$1.19 (1.69) & 2,416,502 \\[3pt]
\midrule
Citation rate before August 31 & 42.57 & 43.45 & 43.89 & 46.11 & \\
\bottomrule
\end{tabular}
\tnote{Percentage points; standard errors clustered by domain in parentheses. *** $p<0.01$, ** $p<0.05$, * $p<0.10$. Estimates of $\beta_1$ and $\beta_2$ in equation~(\ref{eq:page}). Fourth row: the Google-Extended columns refer to hand-written blocks; Cloudflare-managed blocks have a separate term. Panel B is a placebo: the change from the earlier to the later of the two pages before August 31. Last row: each group's citation rate before August 31 in the estimation sample, in percent, with the column's weights; for non-blockers it is 47.78 and 51.86.}
\end{table}

Table~\ref{tab:main} reports the estimates. With the cap (first row), Google-Extended blockers were cited 1.10 points more often after August 31 (standard error 0.33) and other-crawler blockers 0.55 points more (0.41), relative to non-blockers. Without the cap, the estimates are negative and significant, at $-2.60$ and $-1.05$, which suggests that the largest sites drive them. The second row supports this: once the ten largest sites are removed, the uncapped estimates are small and insignificant. Among publishers and news sites alone (third row), the estimate for Google-Extended blockers is about 1.00 under both weightings and significant at the 10 percent level or better, and the estimate for other-crawler blockers is insignificant. A block is the clearest signal when a publisher wrote it by hand: for the 3,014 publishers and news sites with a hand-written Google-Extended block (fourth row), the capped estimate is 0.24 (0.66).

Was the difference there before the change? In part under the capped weighting, and hardly at all under the uncapped one. Panel B of the table runs the same comparison between the two pages before August 31. With the cap, Google-Extended blockers gained 0.60 points on non-blockers (0.36), so they were already drifting up slightly, although the difference is only marginally significant. Giving each group its own linear trend lowers the main estimate to 0.30 (0.59) (Online Appendix Table~\ref{tab:robust}, panel A). Without the cap, the placebo estimate is close to zero for Google-Extended blockers and negative with a large standard error for other-crawler blockers.

\emph{Robustness.} The capped estimate for Google-Extended blockers, 1.10 (0.33), is stable across the checks in Online Appendix Table~\ref{tab:robust}. Changing the cap from 50 to between 10 and 200 observations per domain gives 0.87 to 1.13, and weighting every domain equally gives 1.11 (0.50) (panel C). Clustering by keyword as well as by domain raises the standard error to 0.39 (panel B). Dropping UK sites, some of which had the setting from June 17, gives 1.15 (0.34) (panel A). Adding sites ranked 11 to 20, which are smaller and cited less often, gives 0.30 (0.22) (Online Appendix Table~\ref{tab:top20}). Finally, the sample requires an overview and a top-ten rank after August 31, which could hide withdrawals; two checks on keywords and site-keyword pairs chosen before the change find no sign of this (Online Appendix B.4).

\begin{figure}[htbp]
\centering
\includegraphics[width=\textwidth]{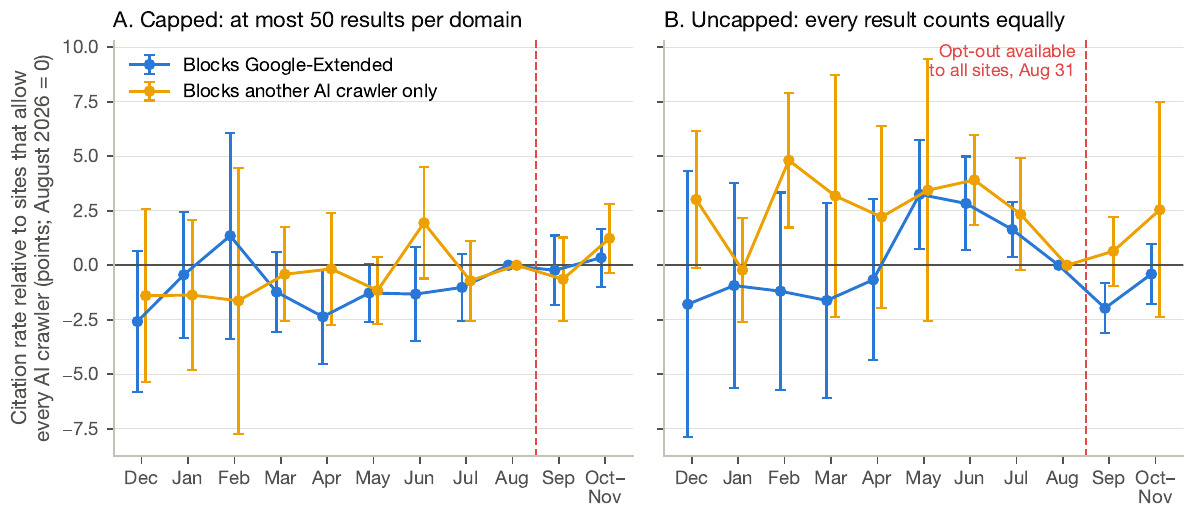}
\caption{Citation rates of blocking sites relative to non-blocking sites, by month}\label{fig:months}
\tnote{Coefficients on group-by-month indicators from equation~(\ref{eq:page}), with the month of capture in place of After; August 2026 is the reference month. Bars are 95 percent confidence intervals. The months hold different numbers of pages (Online Appendix Table~\ref{tab:months}).}
\end{figure}

Figure~\ref{fig:months} shows when any change happened, by replacing After with the month in which each page was captured; each point compares that month with August 2026. In the capped panel, Google-Extended blockers are 0.23 points below August in September (0.82) and 0.35 above it in November (0.68). Most earlier months are 1 to 2.6 points below August, which is why the table's estimate, whose baseline is both pages before, is positive. In the uncapped panel a decline of 1.97 points in September (0.59) is gone by November ($-0.40$, 0.70).

\subsection{The largest sites}\label{sec:largest}

\begin{figure}[htbp]
\centering
\includegraphics[width=\textwidth]{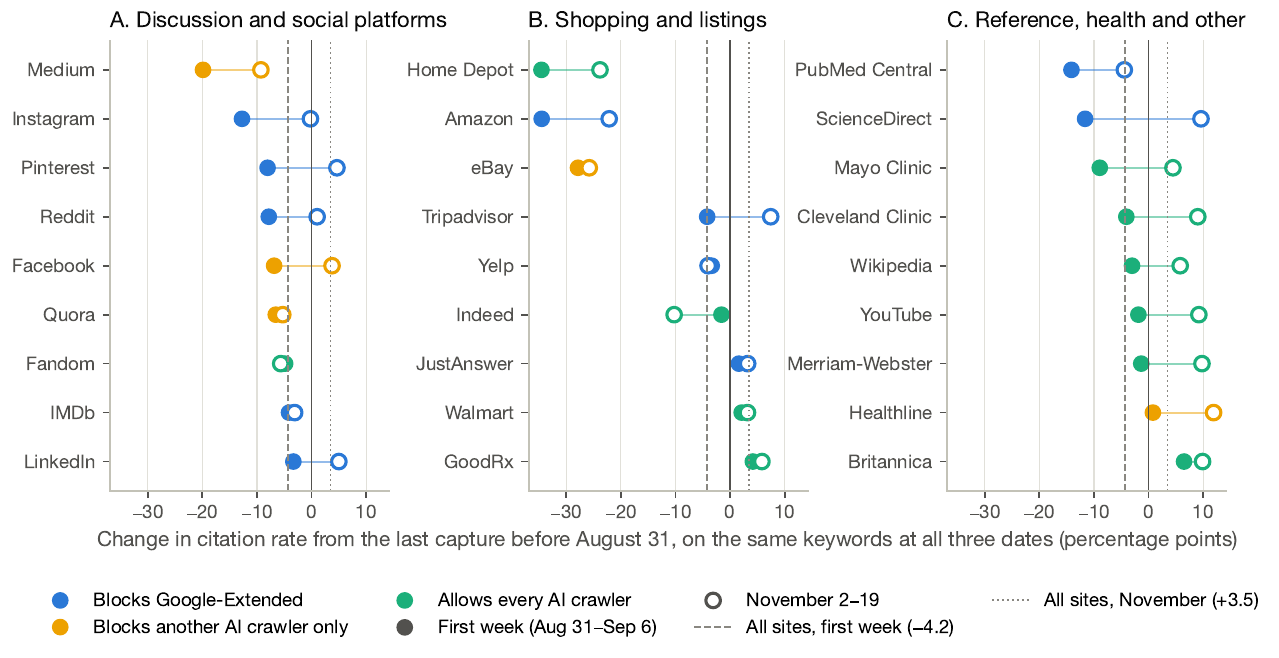}
\caption{Change in citation rate for the largest sites, on the same keywords at three dates}\label{fig:movers}
\tnote{The 11,751 keywords captured in the first week after August 31 and again between November 2 and 19. For each site, the keywords on which it ranks in the top ten on all three pages (62 to 4,596 per site), and the change in the share on which it is cited. Each panel shows the nine sites with the most such keywords.}
\end{figure}

Table~\ref{tab:main} shows that the negative uncapped estimate comes from the largest sites in the sample. Three blockers among them account for it: without Amazon, PubMed Central and Reddit, the uncapped estimate is $-0.23$ (0.51). Did these sites withdraw? Figure~\ref{fig:movers} follows the largest sites on the same keywords. For each site it plots two changes in the citation rate, both measured from the last page before August 31: the change by the first week after, and the change by November. Using the same keywords each time means that no change can come from which keywords were captured. First, the whole market moved: across all sites the citation rate fell 4.2 points in the first week and stood 3.5 points above its earlier level in November. Second, the decline follows the kind of site, not the block. Among discussion platforms, Reddit, which blocks Google-Extended, fell from 51 to 43 percent; Quora and Medium, which do not, fell as much or more. Among shopping sites, Amazon fell from 60 to 25 percent, and eBay and Home Depot, which belong to the other two groups, fell as far. Third, most of the large blockers were back in November on these same keywords: Reddit at 52 percent, and PubMed Central at 53 against 57 before. Amazon was at 37, still 22 points below.

A sustained opt-out of a whole site would take it to zero and keep it there. What we see is a partial decline, shared by blockers and non-blockers of the same kind and mostly reversed by November, which points to a change in how Google chose sources in September. Amazon, PubMed Central and Reddit are all still cited, so none withdrew its whole site; we cannot rule out that one of them excluded a section.

\section{Concluding Remarks}\label{sec:discussion}

Until 2026, a site could leave Google's AI Overviews only at a cost in ordinary Search. When Google removed that cost, the sites with the clearest record of refusing AI stayed: eleven weeks later the share of blocking sites that go uncited has not risen, and the typical blocking site is cited as often as before. The search penalty was therefore not what kept these sites in the overview.

Why did sites stay? Several reasons fit, and our tests do not tell them apart. A cited source receives some clicks through the overview's links; an excluded one keeps only its ordinary listing below it. A site may be waiting for evidence, or for others to move first. Leaving alone achieves little, because the overview continues with other sources. And some sites are paid: according to trade press reports in September 2026, Google pays about a hundred invited publishers when their content contributes to an AI answer. Their names are not public, and they cannot account for the thousands of smaller blockers.

We should caution that our findings face limitations. First, this is entirely a short-run observation, so we cannot speak to exit patterns in the longer run. Second, we only observe AI Overview citations, not the setting itself, with one confirmed case as the only direct validation. 

Our results do not show that the tie was harmless. It could have hurt publishers in two ways that our test cannot detect, because neither makes a site leave. The first is payment. A publisher is paid for its content only if it can threaten to withhold it. Under the tie, withholding meant losing search traffic, so the threat was empty and Google had no reason to pay. The opt-out makes the threat credible, but its value varies. It is worth little to a site whose place another source would take, and more to a publisher that is hard to replace or to publishers acting together, who may now negotiate payment without ever leaving. We do not observe payments. The second is coordination. Each publisher gains from being cited, yet publishers as a group may lose, because the overview answers the query and fewer users click through to any site. No single site can change this by leaving.

Two developments will test these conclusions. First, the CMA requires Google to offer page-level control by March 2027. The current setting covers a whole domain, subdomain or directory; if sites withhold individual pages after declining that coarser choice, then the form of the choice, not a lack of interest, explains the low take-up we find. Second, a right to leave can matter without being used. A publisher that can credibly threaten to withdraw may negotiate better terms, and the licensing deals signed in the coming years will show whether it does.

\clearpage
\setcounter{footnote}{0}
\begin{center}
{\LARGE Online Appendix}\\[6pt]
{\large ``Free to Leave, Choosing to Stay: Websites and Google's AI Overviews Opt-Out''}
\end{center}
\vspace{6pt}

\noindent This appendix has four parts. Part A describes the samples and gives the counts behind the figures. Part B reports additional results. Part C describes the first-week records and the one confirmed opt-out. Part D compares the speed at which sites adopted two earlier controls. ``Capped'' and ``uncapped'' refer to the two weightings of Section~\ref{sec:did}.

\newapp{A}
\section*{A. Samples and counts}

\subsection*{A.1 From keywords to pages}

Serpstat recorded an AI Overview on 10.1 million US keywords in August or September 2026. We drew two million of them: the million with the highest search volume and a random million of the other 9.1 million. We put the two million in random order and retrieved a year of history from DataForSEO for the first 1,045,665, as many as the data budget allowed. Of these, 546,901 keywords have at least one archived capture, and 447,147 have at least one \emph{page}, a capture on which an AI Overview lists a source. The paper uses the 129,057 keywords with at least two pages before August 31, 2026 and at least one after: 30,268 from the high-volume million and 98,789 from the random million. Each contributes its last two pages before August 31 and its first page after. The two pages before are a median of 73 days apart (quartiles 59 and 76), and the page after follows the later of them by a median of 100 days (66 and 107).

The keywords were selected on having an overview in August or September, and a page must have an overview. Section B.4 examines what this selection could hide.

\subsection*{A.2 From sites to each analysis}

Table~\ref{tab:flow} follows the sites through the analyses. The sample has 21,889 blockers (14,638 that block Google-Extended and 7,251 that block another AI crawler only). Three numbers in the paper are subsets of them.

\begin{itemize}\setlength{\itemsep}{1pt}
\item \emph{14,183 blockers identify the regression.} Equation~(\ref{eq:page}) needs a site to rank in the top ten on the same keyword on at least two of the keyword's three pages (15,847 blockers), and the terms of interest need one of those pages to be before August 31 and one after (9,295 + 4,888 = 14,183).
\item \emph{423 blockers are followed one at a time (Section B.6).} They rank on at least 20 pages before August 31 and at least 20 between October 5 and November 19, and were cited on at least a fifth of the earlier pages (240 + 183). The cohort for August 31 to October 4 is built the same way and has 530 blockers (297 + 233). The two cohorts overlap but are not the same sites.
\item \emph{14,871 blockers have a full first-week record.} These come from a different source. Serpstat keeps the 50 most recent citations of each site. We match its sites to the robots.txt classes of every site in the DataForSEO history, not only those in the 129,057 keywords; 9,599 Google-Extended blockers and 5,272 other-crawler blockers have a full record of 50, and 12,366 of them are also in the sample.
\end{itemize}

\begin{table}[htbp]
\centering\footnotesize\setlength{\tabcolsep}{4pt}
\caption{Sample flow}\label{tab:flow}
\begin{tabular}{@{}lrrrrr@{}}
\toprule
 & \multicolumn{3}{c}{Blocks Google-Extended} & Blocks another & Blocks \\
\cmidrule(lr){2-4}
 & All & Hand-written & Cloudflare & crawler only & none \\
\midrule
Sites in the sample & 14,638 & 5,440 & 9,198 & 7,251 & 180,446 \\
\quad of which publishers and news sites & 7,547 & 3,014 & 4,533 & 3,719 & 78,654 \\
Observations (site on a page) & 622,049 & 546,498 & 75,551 & 403,882 & 1,965,774 \\
Sites in the regression sample & 10,463 & 4,040 & 6,423 & 5,384 & 127,586 \\
Sites ranked on the same keyword before and after & 9,295 & 3,650 & 5,645 & 4,888 & 112,761 \\
Followed singly, August 31 to October 4 & 297 & 219 & 78 & 233 & 1,476 \\
Followed singly, October 5 to November 19 & 240 & 181 & 59 & 183 & 1,085 \\
Serpstat first-week records: sites with a full record & 9,599 & -- & -- & 5,272 & 113,812 \\
\quad of which in the sample & 7,927 & 3,358 & 4,569 & 4,439 & 89,805 \\
\bottomrule
\end{tabular}
\tnote{A site is a domain or a subdomain with its own robots.txt file. The sample is the 129,057 keywords with two pages before August 31, 2026 and one after; sites ranked in the top ten organic results. Regression sample: sites that rank on the same keyword on at least two of its three pages. Followed singly: ranked on at least 20 pages before August 31 and at least 20 in the period named, and cited on at least 20 percent of the earlier pages. Sites without an archived robots.txt file (16 percent of observations) are not shown.}
\end{table}

\subsection*{A.3 Counts behind Figure~\ref{fig:exits}, panels B to D}

Table~\ref{tab:bins} gives the numbers plotted in panels B to D of Figure~\ref{fig:exits} and the number of sites behind each point. In every group the lines for the four periods lie close together, and the share of sites not cited, at each number of pages ranked, shows no upward shift after August 31.

\begin{table}[htbp]
\centering\footnotesize
\caption{Share of sites not cited in a period, percent (number of sites in parentheses)}\label{tab:bins}
\begin{tabular}{@{}lcccc@{}}
\toprule
Pages on which the site & June 22 to & July 27 to & August 31 to & October 5 to \\
ranks in the period & July 26 & August 30 & October 4 & November 19 \\
\midrule
\multicolumn{5}{@{}l}{\emph{Google-Extended blockers (Figure 1, panel B)}} \\
1 page & 63.7 (3,964) & 61.6 (3,837) & 61.7 (4,465) & 61.5 (3,929) \\
2 pages & 40.3 (1,195) & 39.2 (1,167) & 40.2 (1,480) & 37.4 (1,231) \\
3 to 4 & 27.4 (804) & 26.4 (759) & 24.9 (1,068) & 21.2 (756) \\
5 to 9 & 10.2 (482) & 13.3 (458) & 14.2 (628) & 7.3 (532) \\
10 to 19 & 4.6 (219) & 5.6 (251) & 5.8 (327) & 2.0 (244) \\
20 or more & 1.3 (234) & 0.4 (240) & 2.4 (330) & 0.4 (259) \\[3pt]
\multicolumn{5}{@{}l}{\emph{Other-crawler blockers (panel C)}} \\
1 page & 61.3 (1,920) & 60.1 (1,845) & 63.6 (2,142) & 61.5 (1,889) \\
2 pages & 43.7 (638) & 39.9 (624) & 41.8 (795) & 36.5 (635) \\
3 to 4 & 24.0 (488) & 23.3 (467) & 25.4 (626) & 20.5 (482) \\
5 to 9 & 12.6 (342) & 13.1 (321) & 12.7 (440) & 10.7 (335) \\
10 to 19 & 5.6 (179) & 5.4 (186) & 6.1 (230) & 2.6 (191) \\
20 or more & 3.4 (177) & 3.7 (190) & 1.6 (257) & 0.5 (202) \\[3pt]
\multicolumn{5}{@{}l}{\emph{Non-blockers (panel D)}} \\
1 page & 58.5 (50,874) & 56.9 (49,376) & 58.3 (59,631) & 57.6 (50,648) \\
2 pages & 35.9 (12,882) & 34.4 (12,796) & 35.9 (17,594) & 32.7 (13,265) \\
3 to 4 & 20.3 (7,566) & 19.7 (7,486) & 22.3 (11,287) & 17.0 (7,892) \\
5 to 9 & 8.4 (3,841) & 8.2 (4,011) & 9.7 (5,893) & 6.6 (4,331) \\
10 to 19 & 3.4 (1,439) & 3.0 (1,470) & 2.8 (2,314) & 2.0 (1,655) \\
20 or more & 2.3 (986) & 2.5 (1,016) & 1.8 (1,585) & 1.5 (1,168) \\[3pt]
\bottomrule
\end{tabular}
\tnote{Sites that rank in the top ten on at least one page in the period. The four periods have 47,505, 49,863, 70,096 and 58,950 pages; in practice the third runs from August 31 to September 13 and the fourth from November 2 to 19.}
\end{table}

\subsection*{A.4 Pages by month}

Figure~\ref{fig:months} plots estimates by month of capture. Each keyword contributes three pages, whose dates depend on when DataForSEO captured it, so the months are not a balanced panel. Table~\ref{tab:months} gives the size of each month. February has the fewest pages and the widest intervals in Figure~\ref{fig:months}.

\begin{table}[htbp]
\centering\footnotesize
\caption{Pages and observations by month of capture}\label{tab:months}
\begin{tabular}{@{}lrrrrr@{}}
\toprule
 & & & Observations in & \multicolumn{2}{c}{Blockers in the regression sample} \\
\cmidrule(lr){5-6}
Month of capture & Pages & Observations & the regression & Google-Extended & Other crawler \\
\midrule
November and December 2025 & 5,882 & 44,681 & 33,302 & 1,259 & 784 \\
January 2026 & 8,242 & 63,881 & 48,547 & 1,567 & 1,000 \\
February & 2,283 & 17,497 & 13,415 & 642 & 403 \\
March & 36,093 & 278,514 & 199,880 & 4,022 & 2,272 \\
April & 20,916 & 161,717 & 123,269 & 3,167 & 1,854 \\
May & 77,475 & 599,679 & 495,357 & 7,250 & 3,908 \\
June & 19,968 & 157,565 & 133,434 & 3,124 & 1,873 \\
July & 54,069 & 420,488 & 364,609 & 6,508 & 3,487 \\
August 1 to 30 & 33,169 & 257,048 & 219,210 & 4,593 & 2,624 \\
August 31 to September 30 & 69,989 & 548,368 & 445,030 & 6,971 & 3,817 \\
October and November & 59,057 & 442,267 & 340,449 & 5,593 & 3,074 \\
\bottomrule
\end{tabular}
\tnote{Observations are sites with a robots.txt class ranked in the top ten on a page. The last two columns count the blocking sites observed in the month. August 1 to 30 is the reference month in Figure~\ref{fig:months}.}
\end{table}

\newapp{B}
\section*{B. Additional results}

\subsection*{B.1 Robustness of the main estimate}

Table~\ref{tab:robust} varies the first row of Table~\ref{tab:main}.

\emph{Specification and sample (panel A).} Leaving out the 4,618 UK sites, 358 of which block Google-Extended, changes nothing. With a separate linear time trend for each blocking group, the capped estimate is 0.30 (0.59), with a 95 percent interval from $-0.86$ to 1.46; the trend itself is 0.64 points per 100 days (0.39). This is the most cautious version, because it attributes the blockers' slow relative rise to a trend that would have continued. Letting the rank effects change after August 31, or giving each sector its own change after August 31, moves the estimates little. The last row follows site-keyword pairs chosen before the change (Section B.4).

\emph{Clustering (panel B).} Clustering by keyword as well as by domain raises the capped standard error from 0.33 to 0.39 and leaves the uncapped one unchanged.

\emph{Weights (panel C).} The capped weighting gives a domain with $n$ observations a total weight of $\min(n, 50)$. Caps of 10 to 200 give estimates between 0.87 and 1.13. Giving every domain a total weight of one gives 1.11 (0.50).

\emph{Kind of block (panel D).} Hand-written blocks are a more deliberate statement than Cloudflare's one-click setting. For hand-written blocks the capped estimate is 0.86 (0.45) over all sites and 0.24 (0.66) among publishers and news sites. The uncapped estimate for hand-written blocks, $-3.08$ (1.02), is where the three very large sites of Section~\ref{sec:largest} sit.

\emph{Absolute changes.} On the site-keyword pairs seen before and after August 31, with the capped weights, the citation rate went from 45.0 to 44.4 percent for Google-Extended blockers, from 46.0 to 44.3 for other-crawler blockers, and from 50.3 to 48.1 for non-blockers. Uncapped, the three figures are 46.0 to 42.3, 48.1 to 45.9 and 54.4 to 52.9.

\begin{table}[htbp]
\centering\scriptsize\setlength{\tabcolsep}{2pt}
\caption{Robustness of the change in citation rate after August 31}\label{tab:robust}
\begin{tabular}{@{}lccccr@{}}
\toprule
 & \multicolumn{2}{c}{Capped at 50 results per domain} & \multicolumn{2}{c}{Uncapped} & \\
\cmidrule(lr){2-3}\cmidrule(lr){4-5}
 & Blocks Google- & Blocks another & Blocks Google- & Blocks another & \\
 & Extended & AI crawler only & Extended & AI crawler only & Observations \\
\midrule
\multicolumn{6}{@{}l}{\emph{A. Changes to the specification or the sample}} \\
Table 2, first row & 1.10*** (0.33) & 0.55 (0.41) & $-$2.60*** (0.91) & $-$1.05** (0.49) & 2,416,502 \\
Without UK sites & 1.15*** (0.34) & 0.56 (0.42) & $-$2.58*** (0.92) & $-$1.10** (0.49) & 2,372,156 \\
Linear trend for each blocking group & 0.30 (0.59) & $-$0.73 (0.70) & $-$4.02*** (1.32) & $-$0.44 (1.18) & 2,416,502 \\
Rank effects change after August 31 & 1.00*** (0.33) & 0.45 (0.41) & $-$2.46*** (0.93) & $-$1.37*** (0.51) & 2,416,502 \\
Sector-specific change after August 31 & 1.03*** (0.34) & 0.49 (0.41) & $-$2.16** (0.88) & $-$1.12* (0.62) & 2,416,502 \\
Pairs chosen before August 31 (Section B.4) & 0.45 (0.47) & 0.01 (0.57) & $-$2.21** (0.93) & 1.71 (2.24) & 2,001,070 \\[3pt]
\multicolumn{6}{@{}l}{\emph{B. Standard errors clustered by domain and by keyword}} \\
Table 2, first row & 1.10*** (0.39) & 0.55 (0.47) & $-$2.60*** (0.91) & $-$1.05** (0.50) & 2,416,502 \\[3pt]
\multicolumn{6}{@{}l}{\emph{C. Other weights (in place of the cap of 50)}} \\
Cap of 10 results per domain & 1.13*** (0.40) & 0.10 (0.49) & & & 2,416,502 \\
Cap of 20 & 1.13*** (0.37) & 0.23 (0.45) & & & 2,416,502 \\
Cap of 100 & 0.99*** (0.32) & 0.68* (0.39) & & & 2,416,502 \\
Cap of 200 & 0.87*** (0.31) & 0.68* (0.38) & & & 2,416,502 \\
Every domain weighted equally & 1.11** (0.50) & $-$0.05 (0.62) & & & 2,416,502 \\[3pt]
\multicolumn{6}{@{}l}{\emph{D. Google-Extended blockers by kind of block}} \\
All sites, hand-written & 0.86* (0.45) & 0.55 (0.41) & $-$3.08*** (1.02) & $-$1.07** (0.49) & 2,416,502 \\
All sites, Cloudflare-managed & 1.35*** (0.47) & & 1.40** (0.56) & & \\
Publishers and news, hand-written & 0.24 (0.66) & $-$0.32 (0.63) & 0.74 (0.50) & 0.73 (0.46) & 513,017 \\
Publishers and news, Cloudflare-managed & 1.90** (0.75) & & 1.73** (0.69) & & \\
\bottomrule
\end{tabular}
\tnote{Percentage points; standard errors in parentheses, clustered by domain except in panel B. *** $p<0.01$, ** $p<0.05$, * $p<0.10$. Each row is equation~(\ref{eq:page}) with the change named, except the last row of panel A, which is described in Section B.4. In panel D each pair of rows is one regression with separate After terms for hand-written and Cloudflare-managed Google-Extended blocks.}
\end{table}

\subsection*{B.2 Sites ranked 11 to 20}

The main text follows sites ranked in the top ten organic results. Extending the sample to the top 20 does not change the conclusion.

The archived pages record results below the tenth for 99 percent of captures and down to the twentieth for 74 percent. Ranks 11 to 20 bring in smaller sites. The number of Google-Extended blockers rises from 14,638 to 20,650, and the number that rank on the same keyword before and after August 31 from 9,295 to 13,509. Two things are weaker at these ranks. The sites are cited much less often: Google-Extended blockers were cited on 12 percent of pages before August 31, against 43 percent in the top ten. And fewer of them have a robots.txt file on record: 78 percent of observations have a class, against 84 percent in the top ten.

Table~\ref{tab:top20} reports equation~(\ref{eq:page}) on the larger sample. With the cap, Google-Extended blockers were cited 0.30 points more often after August 31 (standard error 0.22), against 1.10 (0.33) in the top ten. Their citation rate before was 28 percent, so an opt-out by 3 percent of them would have lowered it by about 0.85 points. The 95 percent interval runs from $-0.14$ to 0.73. At ranks 11 to 20 alone the estimate is 0.19 (0.28).

Uncapped, the estimate is $-1.34$ (0.77), and $-0.33$ (0.49) without the ten largest sites. As in the main text, the negative estimate comes from the largest sites. It is absent at ranks 11 to 20, where the estimate is 0.56 (0.48).

\begin{table}[htbp]
\centering\footnotesize\setlength{\tabcolsep}{3.5pt}
\caption{Change in citation rate after August 31, with sites ranked 11 to 20}\label{tab:top20}
\begin{tabular}{@{}lccccr@{}}
\toprule
 & \multicolumn{2}{c}{Capped at 50 results per domain} & \multicolumn{2}{c}{Uncapped} & \\
\cmidrule(lr){2-3}\cmidrule(lr){4-5}
 & Blocks Google- & Blocks another & Blocks Google- & Blocks another & \\
 & Extended & AI crawler only & Extended & AI crawler only & Observations \\
\midrule
\multicolumn{6}{@{}l}{\emph{A. After August 31}} \\
Ranks 1 to 10 (Table 2) & 1.10*** (0.33) & 0.55 (0.41) & $-$2.60*** (0.91) & $-$1.05** (0.49) & 2,416,502 \\
Ranks 1 to 20 & 0.30 (0.22) & 0.22 (0.27) & $-$1.34* (0.77) & $-$0.26 (0.44) & 4,356,503 \\
\quad without the 10 largest sites & 0.31 (0.22) & 0.23 (0.27) & $-$0.33 (0.49) & $-$0.17 (0.51) & 3,280,406 \\
\quad publishers and news sites & 0.28 (0.32) & $-$0.04 (0.40) & $-$0.06 (0.33) & 0.55 (0.43) & 1,146,228 \\
Ranks 11 to 20 only & 0.19 (0.28) & 0.21 (0.33) & 0.56 (0.48) & 0.97 (0.60) & 1,194,417 \\[3pt]
\multicolumn{6}{@{}l}{\emph{B. Citation rate before August 31}} \\
Ranks 1 to 10 (Table 2) & 42.57 & 43.45 & 43.89 & 46.11 & \\
Ranks 1 to 20 & 28.23 & 28.43 & 32.45 & 31.99 & \\
Ranks 11 to 20 only & 11.75 & 11.96 & 12.97 & 14.03 & \\
\bottomrule
\end{tabular}
\tnote{Percentage points; standard errors clustered by domain in parentheses. *** $p<0.01$, ** $p<0.05$, * $p<0.10$. Estimates of $\beta_1$ and $\beta_2$ in equation~(\ref{eq:page}), with one rank effect for each position. The first row repeats the first row of Table~\ref{tab:main}. The pages and the site classes are those of the main text; a site's rank on a page is its best organic position. The indented rows are subsets of the second row. The first and last rows do not add up to the second because a site that ranks on a keyword in the top ten on one page and below it on another is followed only in the second. The ten largest sites are those with the most observations at ranks 1 to 20. Panel B gives each group's citation rate before August 31 in the estimation sample, with the column's weights; for sites that block no AI crawler it is 33.15 and 38.20 at ranks 1 to 20.}
\end{table}

\subsection*{B.3 Listings without a text snippet}

Table~\ref{tab:controls} says that the snippet tags keep a page out of the overview. Our data agree, and they show that almost no site used them.

Each archived results page records the text snippet under every organic result. A result with none is how a page with \texttt{nosnippet} or \texttt{max-snippet:0} appears. It is also how a page that Googlebot may not crawl appears, so the measure covers both controls in Table~\ref{tab:controls} that cost search traffic and cannot tell them apart. We leave out video results, which never carry a snippet (4.6 percent of results), and captures before February 2026.\footnote{In December 2025 and January 2026, 3 percent of results have no snippet, against 0.2 percent in each later month, which points to how those captures were recorded.}

Listings without a snippet are rare: 0.2 percent of 3.3 million results. The share is 0.19, 0.22 and 0.19 percent on the two pages before August 31 and the page after, so it did not change when the new setting arrived.

{\sloppy Sites whose listings have no snippet are almost never cited. Of the 15,721 sites with at least 20 results, three have no snippet on 90 percent or more of them (fueleconomy.gov, leginfo.legislature.ca.gov and m.imdb.com). They are cited on 4 of their 309 results. The 15,129 sites that always show a snippet are cited on 46 percent of theirs. Result by result the measure is noisier. Results without a snippet are cited 14 percent of the time and results with one 44 percent, which suggests that a snippet is sometimes missing for other reasons.\par}

Two sites show that the measure separates the two ways out. Reuters, the one large blocker that fell to zero (Section B.6), shows a snippet on all 558 of its listings, including the 108 after October 5 on which it is never cited. That fits the new setting and not the snippet tags. TreasuryDirect went the other way. Before August 31, 2 of its 26 listings had no snippet and 19 were cited. After, all 12 had none and none was cited. It dropped out of the overview together with its snippets, which is the mark of the older controls. Of the 9,995 sites with at least 10 results on each side of August 31, three moved from under 10 percent of listings without a snippet to half or more (TreasuryDirect, cheese.com and nass.usda.gov), and none moved the other way.

\subsection*{B.4 Does the way the sample was chosen hide withdrawals?}

Three features of the sample condition on outcomes after August 31. The keywords had an overview in August or September. A page must have an overview. And the regression follows a site only where it ranks in the top ten on a keyword both before and after. A withdrawal could in principle remove an overview, or coincide with a loss of rank, and so drop out of the sample. Two checks use cohorts fixed before the change.

\emph{Is there still an overview?} Take every keyword with at least two pages before August 31 (253,820 keywords), and look at its next capture, whether or not it has an overview. About half had been captured again by November 19. Table~\ref{tab:select} shows the share of those captures with an overview that lists sources. It is 96.6 percent where the last page before had cited a Google-Extended blocker and 96.1 percent where no such blocker ranked. The same calculation one step earlier, for keywords chosen before June 22, gives 93.7 and 94.1 percent. Overviews did not disappear where blockers had been cited.

\begin{table}[htbp]
\centering\footnotesize
\caption{Is there an overview at the keyword's next capture?}\label{tab:select}
\begin{tabular}{@{}lrcc@{}}
\toprule
 & & Captured again & Next capture has an overview \\
On the last page before, the overview & Keywords & in the window (\%) & that lists sources (\%) \\
\midrule
\multicolumn{4}{@{}l}{\emph{A. Keywords with two or more pages before August 31; next capture August 31 to November 19}} \\
Cites a Google-Extended blocker & 165,122 & 53.4 & 96.6 \\
Ranks one in the top ten, cites none & 60,465 & 51.1 & 96.5 \\
Neither & 28,233 & 50.2 & 96.1 \\[3pt]
\multicolumn{4}{@{}l}{\emph{B. Keywords with two or more pages before June 22; next capture June 22 to August 30}} \\
Cites a Google-Extended blocker & 150,038 & 74.0 & 93.7 \\
Ranks one in the top ten, cites none & 44,610 & 73.1 & 93.8 \\
Neither & 23,569 & 70.6 & 94.1 \\[3pt]
\bottomrule
\end{tabular}
\tnote{All keywords with at least two overview pages before the date, grouped by the last of those pages. The next capture is the first archived capture in the window, with or without an AI Overview.}
\end{table}

\emph{Does the estimate depend on ranking afterwards?} Take every site-keyword pair in which the site ranks in the top ten on the last page before August 31 (1.0 million pairs), and ask whether the site is cited on the keyword's page after, at any rank or outside the ranked results. The outcome is the change in whether the site is cited. The comparison is with the same construction one step earlier, from the earlier page before to the later one. Relative to non-blockers and to that earlier step, the change for Google-Extended blockers is 0.45 points (0.47) capped and $-2.21$ (0.93) uncapped (Table~\ref{tab:robust}, panel A, last row). These are close to the estimates in Table~\ref{tab:main}, so conditioning on rank after August 31 does not drive them. As for rank itself, 74.0 percent of Google-Extended blockers' pairs are still in the top ten on the page after, against 75.4 percent one step earlier; for non-blockers the figures are 70.2 and 69.6.

\subsection*{B.5 UK sites and the early rollout}

Google honored the setting for a subset of UK site owners from June 17, 2026, and opened it to all websites worldwide on August 31. Google has not said how the UK subset was defined, or whether an exclusion set by a UK owner applied to results shown outside the UK. We therefore treat UK sites as possibly treated early. A site counts as UK if its domain is under \dom{.uk} or if it is on a list of 71 sites of UK-headquartered publishers that use another top-level domain (the BBC, the Guardian, the Financial Times, Future plc's titles and others). The list is ours and is not exhaustive.

In the sample, 4,618 sites are UK sites; 358 block Google-Extended and 208 block another AI crawler. They hold 1.9 percent of observations. Without them the capped estimate is 1.15 (0.34) and the uncapped one $-2.58$ (0.92) (Table~\ref{tab:robust}). For UK Google-Extended blockers alone the capped estimate is 4.25 (2.13).

An early opt-out would show between June 17 and August 30. Using all archived pages, 54 UK blockers ranked on at least 20 pages before June 17 and at least 20 between June 17 and August 30, and were cited on at least a fifth of the earlier ones. None was at zero in the later window, and the median site's citation rate was unchanged (ratios of 1.06 for the 19 Google-Extended blockers and 1.00 for the 35 other-crawler blockers). Panel B of Table~\ref{tab:main} is the corresponding test for all sites.

\subsection*{B.6 Large sites followed one at a time}

\begin{figure}[htbp]
\centering
\includegraphics[width=0.55\textwidth]{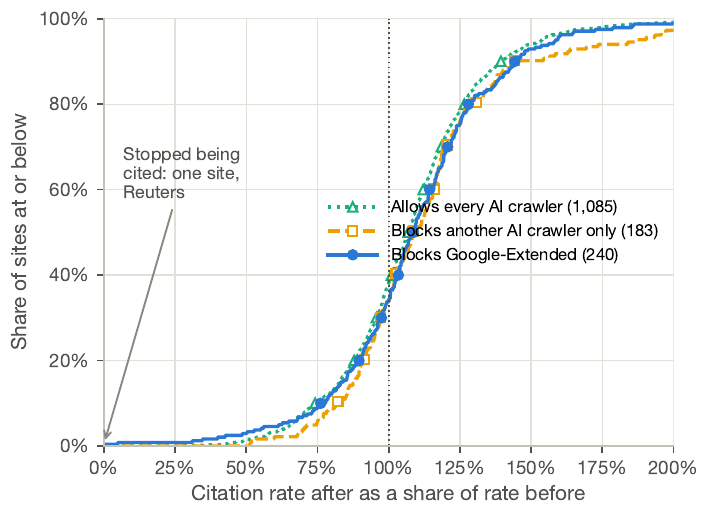}
\caption{Sites followed one at a time: citation rate after as a share of the rate before}\label{fig:tracked}
\tnote{Sites that ranked on at least 20 pages before August 31 and at least 20 between October 5 and November 19, and were cited on at least a fifth of the earlier pages. Cumulative distribution of each site's later citation rate as a share of its rate before, with a marker every ten percentiles.}
\end{figure}

Some sites rank on enough pages to be followed one at a time: those that ranked on at least 20 pages before August 31 and at least 20 between October 5 and November 19, and were cited on at least a fifth of the earlier pages. They are 240 Google-Extended blockers, 183 other-crawler blockers and 1,085 non-blockers. For each site we divide its citation rate in the later period by its rate before, and Figure~\ref{fig:tracked} plots the distribution of this ratio in each group. A site that withdrew would have a ratio of zero, and if many blockers had withdrawn in part, their line would lie to the left of the non-blockers' line. The three lines coincide, and the median site in each group is cited 7 to 9 percent more often than before. One of the 423 blockers is at zero.

\emph{How often does a site fall to zero without an opt-out?} Table~\ref{tab:track} applies the same rule to three comparisons. After the change, one blocker of 423 is at zero in October and November and none of 530 in September. The one is Reuters, which blocks Google-Extended. It was cited on 37 percent of the pages where it ranked before August 31 and on 50 percent in September. Between November 2 and 19 it ranked on 108 pages and was cited on none, while its listings kept their text snippets (Section B.3). Reuters has not said whether it used the setting; in June the general manager of Reuters Digital said the company would evaluate it, ``probably testing as well.''\footnote{J.~Davies and S.~Guaglione, ``Google's AI opt-out leaves publishers with a choice they can't safely use,'' \emph{Digiday}, June 8, 2026, \url{https://digiday.com/media/googles-ai-opt-out-leaves-publishers-with-a-choice-they-cant-safely-use/} (\accessed).} We cannot tell what happened. Before the change, one blocker of 386 is at zero (\dom{allareacodes.com}, which blocks another AI crawler; it was cited on 40 percent of the pages where it ranked until June, on none of 243 between July 27 and August 30, and again on 56 of 77 in November). With all archived pages in place of the three per keyword, the comparison before the change finds 2 of 1,032 blockers at zero and the comparison after finds 1 of 835, again Reuters.

If Reuters is counted as an opt-out, the share of closely tracked blockers out of the overview in October and November is 1 of 423, or 0.24 percent, with an exact (Clopper--Pearson) 95 percent interval from 0.01 to 1.31 percent. Among Google-Extended blockers alone it is 1 of 240, with an interval from 0.01 to 2.30 percent. For September it is 0 of 530, with an upper limit of 0.69 percent. If the true share were 1 percent, the probability of finding at most one such site among 423 would be 0.08.

\begin{table}[htbp]
\centering\footnotesize\setlength{\tabcolsep}{3.5pt}
\caption{Sites followed one at a time: at zero, and down by half or more}\label{tab:track}
\begin{tabular}{@{}lrrrrrrrrr@{}}
\toprule
 & \multicolumn{3}{c}{Blocks Google-Extended} & \multicolumn{3}{c}{Blocks another crawler} & \multicolumn{3}{c}{Blocks none} \\
\cmidrule(lr){2-4}\cmidrule(lr){5-7}\cmidrule(lr){8-10}
Later period & Sites & Zero & Half & Sites & Zero & Half & Sites & Zero & Half \\
\midrule
Oct.\ 5 to Nov.\ 19 (Figure B1) & 240 & 1 & 7 & 183 & 0 & 0 & 1,085 & 0 & 15 \\
Aug.\ 31 to Oct.\ 4 & 297 & 0 & 12 & 233 & 0 & 8 & 1,476 & 2 & 30 \\
July 27 to Aug.\ 30 (placebo) & 218 & 0 & 5 & 168 & 1 & 4 & 934 & 0 & 23 \\
\bottomrule
\end{tabular}
\tnote{Each row compares a later period with all pages before it: before August 31 in the first two rows, before July 27 in the third, which lies wholly before the change. Sites ranked on at least 20 pages in each period and cited on at least 20 percent of the pages in the earlier one. Zero: not cited on any page in the later period. Half: citation rate in the later period below half the rate in the earlier one (includes the sites at zero).}
\end{table}

\subsection*{B.7 The publishers that sued}

Penske Media and Chegg sued Google in 2025 over AI Overviews. Their sites are the clearest cases of a stated objection. Table~\ref{tab:sued} uses all archived pages. Penske Media's sites were cited on 38 percent of the pages where they ranked in the top ten before August 31, on 35 percent between August 31 and October 4, and on 43 percent between October 5 and November 19. Nine of its sites block Google-Extended. Chegg, which blocks no AI crawler, was cited on 31, 41 and 35 percent. None of these sites has withdrawn its whole site.

\begin{table}[htbp]
\centering\footnotesize
\caption{Citations of the sites of the publishers that sued Google}\label{tab:sued}
\begin{tabular}{@{}llrrrrrr@{}}
\toprule
 & & \multicolumn{2}{c}{Before August 31} & \multicolumn{2}{c}{Aug.\ 31 to Oct.\ 4} & \multicolumn{2}{c}{Oct.\ 5 to Nov.\ 19} \\
\cmidrule(lr){3-4}\cmidrule(lr){5-6}\cmidrule(lr){7-8}
Site & Blocks in August & Pages & Cited (\%) & Pages & Cited (\%) & Pages & Cited (\%) \\
\midrule
\texttt{billboard.com} & Google-Extended & 1,378 & 44 & 152 & 41 & 168 & 48 \\
\texttt{rollingstone.com} & Google-Extended & 1,298 & 40 & 139 & 29 & 122 & 47 \\
\texttt{hollywoodreporter.com} & Google-Extended & 687 & 37 & 81 & 32 & 89 & 42 \\
\texttt{variety.com} & Google-Extended & 656 & 35 & 82 & 44 & 78 & 37 \\
\texttt{deadline.com} & Google-Extended & 348 & 36 & 48 & 35 & 32 & 41 \\
\texttt{robbreport.com} & Google-Extended & 225 & 34 & 21 & 43 & 23 & 26 \\
\texttt{wwd.com} & Google-Extended & 218 & 49 & 27 & 37 & 21 & 43 \\
\texttt{sheknows.com} & Google-Extended & 217 & 33 & 14 & 21 & 31 & 45 \\
\texttt{goldderby.com} & Google-Extended & 193 & 32 & 20 & 25 & 15 & 33 \\
\texttt{indiewire.com} & No file on record & 121 & 29 & 14 & 21 & 13 & 46 \\
\texttt{sportico.com} & No file on record & 121 & 36 & 7 & 71 & 6 & 17 \\
\texttt{bgr.com} & None & 109 & 26 & 16 & 19 & 12 & 42 \\
\texttt{tvline.com} & None & 103 & 35 & 11 & 27 & 18 & 50 \\
All Penske Media sites (19) & & 6,001 & 38 & 662 & 35 & 665 & 43 \\[3pt]
\texttt{chegg.com} & None & 316 & 31 & 37 & 41 & 43 & 35 \\
\bottomrule
\end{tabular}
\tnote{All archived pages with an AI Overview on which the site ranks in the top ten. Penske Media sites with at least 100 pages before August 31 are listed; the total covers 19 of its sites.}
\end{table}

\subsection*{B.8 Citation rates before the opt-out, at the same rank on the same page}

Before August 31, blockers were already cited somewhat less often than non-blockers. On the two pages before August 31, with page effects and rank effects, a Google-Extended blocker is cited 2.73 points less often than a non-blocker at the same rank on the same page (0.25) with the capped weights, and an other-crawler blocker 1.59 points less often (0.33). Uncapped, the differences are 8.50 (2.29) and 1.40 (2.26); without the ten largest sites the uncapped difference for Google-Extended blockers is 2.50 (1.03). The large uncapped gap again reflects a few very large platforms. The site-by-keyword effects in equation~(\ref{eq:page}) absorb these differences in levels.

\newapp{C}
\section*{C. First-week records and the confirmed opt-out}

\subsection*{C.1 The records}

Serpstat's AI Overview Records hold, for each site, its 50 most recent citations in AI Overviews, each with the keyword and a timestamp. Our copy ends on September 7, 2026 at 16:47. A site that is cited often turns its record over within a day or two, so its latest citation is on the last days of the file. A site that stops being cited keeps its record, and its latest citation stays where it was. Sites are not dropped when citations stop: 98,010 sites in the file, mostly small ones with fewer than 50 records, have a latest citation in August.

Panel A of Figure~\ref{fig:exits} uses the sites with a full record of 50. Table~\ref{tab:firstweek} gives the distribution behind it. The test asks whether a site was last cited before the setting took effect. It cannot see a site that left and came back within the week, and it says nothing about sites cited too rarely to fill a record.

Ten blockers have a latest citation before September 3. Four of them are cited again on DataForSEO pages captured later, so they had not withdrawn. The other six rank on three pages or fewer after August 31, too few to tell.

\begin{table}[htbp]
\centering\footnotesize
\caption{Date of the latest citation, sites with a full record of 50}\label{tab:firstweek}
\begin{tabular}{@{}lrrr@{}}
\toprule
Latest citation & Blocks Google-Extended & Blocks another crawler only & Blocks none \\
\midrule
Before August 31 & 2 & 1 & 7 \\
August 31 & 1 & 0 & 6 \\
September 1 & 0 & 0 & 20 \\
September 2 & 6 & 0 & 24 \\
September 3 & 2 & 1 & 36 \\
September 4 & 2 & 5 & 60 \\
September 5 & 30 & 11 & 336 \\
September 6 & 945 & 501 & 12,904 \\
September 7 (records end at 16:47) & 8,611 & 4,753 & 100,419 \\
\midrule
Sites with a full record & 9,599 & 5,272 & 113,812 \\
\bottomrule
\end{tabular}
\end{table}

\subsection*{C.2 Zyppy}

Zyppy, a firm that advises websites on search engine optimization (SEO), turned the setting on for its site on September 1, 2026 and off again on September 29, and published the result.\footnote{C.~Shepard, ``We Opted Out of Google AI Search: Here's What Happened,'' Zyppy, November 5, 2026, \url{https://signal.zyppy.com/p/we-opted-out-of-google-ai-search} (\accessed); B.~Schwartz, Search Engine Roundtable, October 30, 2026.} It reports that its impressions in AI features, as counted by Search Console, fell by more than 99 percent within an hour, that its links disappeared from overviews on queries where it still ranked first, and that both returned when it opted back in. It is the only owner-confirmed case we found; a search of the trade press on November 22, 2026 found no named publisher that has said it used the setting.

Our data, which do not depend on Search Console, show the same break in two sources.

\begin{itemize}\setlength{\itemsep}{1pt}
\item \emph{Serpstat.} Among overview pages on which \dom{zyppy.com} ranked in the top ten, it is cited on 3 of 7 between August 20 and 31 and on none of 6 between September 3 and 6, on one of which it ranked first and on two second. Its latest entry in the AI Overview Records is on September 1.
\item \emph{DataForSEO.} For one keyword it ranked in the top three on four captures. It was cited on all three before August 31 and not on September 10. No capture falls after September 29, so its return is not observed.
\end{itemize}

Pooled, Zyppy is cited on 6 of 10 pages before and on none of 7 during its opt-out (Fisher exact $p=0.02$). This is one small site. It shows that the setting worked there from the first day and that our sources registered it. It does not measure how reliably a withdrawal would be detected across sites, or for a single directory.

\newapp{D}
\section*{D. How fast did sites adopt two earlier controls?}

This part compares the eleven weeks after August 31 with the speed at which sites adopted two earlier controls: blocking GPTBot, possible from August 7, 2023, and blocking Google-Extended, possible from September 28, 2023. Both take one line in a file the site already serves.

We took the 230 most-cited sites in an earlier, Serpstat-based sample (150 that blocked GPTBot or Google-Extended in November 2026 and 80 that blocked neither) and collected every copy of their robots.txt files in the Internet Archive from July 2023 to March 2024, a median of 20 days apart. Of the sites that block the crawler today, 93 (GPTBot) and 87 (Google-Extended) have at least three archived copies in that window. Because copies are spaced, a block is first seen some days after it was added. Table~\ref{tab:react} therefore gives two numbers for each date: the share already seen blocking, which is a lower bound on adoption, and the share not yet seen allowing after that date, which is an upper bound.

\begin{table}[htbp]
\centering\footnotesize
\caption{Share of today's blockers that had added the block, by days since the control became available (percent)}\label{tab:react}
\begin{tabular}{@{}lcccc@{}}
\toprule
 & \multicolumn{2}{c}{GPTBot (93 sites)} & \multicolumn{2}{c}{Google-Extended (87 sites)} \\
\cmidrule(lr){2-3}\cmidrule(lr){4-5}
Days since launch & At least & At most & At least & At most \\
\midrule
7 & 3 & 36 & 3 & 22 \\
14 & 12 & 39 & 8 & 24 \\
30 & 38 & 47 & 20 & 28 \\
60 & 48 & 56 & 22 & 29 \\
90 & 57 & 60 & 25 & 30 \\
120 & 58 & 60 & 29 & 33 \\
\bottomrule
\end{tabular}
\tnote{Sites that block the crawler in November 2026 and have at least three archived copies of robots.txt in the launch window. ``At least'': share seen blocking by that day. ``At most'': share not seen allowing after that day.}
\end{table}

Among sites that block GPTBot today, 38 to 47 percent had added the block within 30 days and 58 to 60 percent within 120 days. Google-Extended was adopted more slowly: 20 to 28 percent within 30 days and 29 to 33 percent within 120. The main text uses the GPTBot pace, the faster of the two, so that the comparison does not understate how much adoption should be visible by now. The comparison is loose in both directions. The earlier controls cost a site nothing, which favors faster adoption then. The new setting was demanded publicly for a year and announced in advance, which favors faster adoption now.

\end{document}